\documentclass[reprint,superscriptaddress,longbibliography,amsmath,amssymb,aps,prl,floatfix]{revtex4-1}

\usepackage{graphicx}% Include figure files
\usepackage{dcolumn}% Align table columns on decimal point
\usepackage{bm}% bold math
\usepackage{hyperref}% add hypertext capabilities
\usepackage{color}

\usepackage{todonotes}
\usepackage{subcaption}
\usepackage{physics}
\newcommand{\ilm}{Univ Lyon, Univ Claude Bernard Lyon 1, CNRS, Institut Lumi\`ere Mati\`ere, F-69622, VILLEURBANNE, France}
\newcommand{\iuf}{Institut Universitaire de France (IUF), 1 rue Descartes, 75005 Paris, France}

\newcommand{\zs}{z_\mathrm{s}}              % HBC
\newcommand{\vs}{v_\mathrm{s}}              % slip velocity
\newcommand{\debye}{\lambda_\text{D}}   % Debye length
\newcommand{\nbulk}{n_\mathrm{s}}
\newcommand{\tosm}{\text{to}}           % thermo-osmosis
\newcommand{\beff}{b_\mathrm{eff}}      % effective slip length = b - \zs
\newcommand{\dl}{d_\ell}                % depletion layer
\newcommand{\sgn}{\text{sgn}\qty}
\newcommand{\eps}{\varepsilon} 
\newcommand{\rhoe}{\rho_\text{e}}
\begin{document}

%\preprint{APS/123-QED}

\title{Fast and Versatile Thermo-osmotic Flows with a Pinch of Salt}

\author{Cecilia Herrero}
\email{cecil.herr@gmail.com}
\affiliation{\ilm}
\author{Michael De San F\'eliciano}
\affiliation{\ilm}
\author{Samy Merabia}
\email{samy.merabia@univ-lyon1.fr}
\affiliation{\ilm}
\author{Laurent Joly}
\email{laurent.joly@univ-lyon1.fr}
\affiliation{\ilm}
\affiliation{\iuf}

\date{\today}% It is always \today, today,

\begin{abstract}
Thermo-osmotic flows -- flows generated in micro and nanofluidic systems by thermal gradients -- could provide an alternative approach to harvest waste heat. However, such use would require massive thermo-osmotic flows, which are up to now only predicted for special and expensive materials. There is thus an urgent need to design affordable nanofluidic systems displaying large thermo-osmotic coefficients. In this paper we propose a general model for thermo-osmosis of aqueous electrolytes in charged nanofluidic channels, taking into account hydrodynamic slip, together with the different solvent and solute contributions to the thermo-osmotic response. We apply this model to a wide range of systems, by studying the effect of wetting, salt type and concentration, and surface charge. We show that intense thermo-osmotic flows can be generated using slipping charged surfaces. We also predict for intermediate wettings a transition from a thermophobic to a thermophilic behavior depending on the surface charge and salt concentration. Overall, this theoretical framework opens an avenue for controlling and manipulating thermally induced flows with common charged surfaces and a pinch of salt.
\end{abstract}

\maketitle

%\section*{Introduction}
Due to the increasing world energy consumption and the need of new clean energies, waste heat harvesting is a major challenge for the decades to come. Some of the most common difficulties to harvest waste heat come from the small temperature differences between the source and the environment ($< 50\,^\circ\,$C) \cite{Straub2016}, as well as from the need to use rare, expensive and often toxic thermoelectric materials \cite{Kristiansen2019}. 
Alternatively, thermo-osmotic flows (generated at liquid-solid interfaces by temperature gradients) can be used to transform waste heat into electricity via a turbine \cite{Straub2017}, or to pump water for desalination \cite{Zhao2015,Oyarzua2017}.
Historically, Derjaguin and Sidorenkov measured the first reported water flow by applying a temperature gradient through porous glass \cite{Derjaguin1941}. Since then, a broad literature has been devoted to the 
measure of the thermo-osmotic response, whether from experiments \cite{Dariel1975,Mengual1978,Piazza2004,Barragan2017} or molecular dynamics (MD) simulations \cite{Ganti2017,Fu2017,Rajegowda2018analysing,Ganti2018,Prakash2020}. Nevertheless, some disagreements have been reported in the results for aqueous electrolytes, with a thermo-osmotic response observed for pure water and uncharged membranes \cite{Mengual1978}, and disagreements in the flow direction (toward the hot side, so-called thermophilic flow, or toward the cold side, so-called thermophobic flow) for similar systems \cite{Derjaguin1980,Rusconi2004,Nedev2015,Bregulla2016}. Such differences cannot be understood by the classical theory \cite{Anderson1989} developed by Derjaguin and Sidorenkov \cite{Derjaguin1941,Derjaguin1987}, and by Ruckenstein for thermophoresis \cite{Ruckenstein1981}. This theory, based on the electrostatic enthalpy of the electric double layer (EDL) appearing close to charged walls \cite{Markovich2016charged}, predicts that the flow is controlled by the electric surface charge, and always goes to the hot side.

Thermo-osmosis has seen a renewed interest due to the massive thermo-osmotic responses predicted by the use of novel materials, such as soft nanochannels \cite{Maheedhara2018}, carbon-nanotubes \cite{Oyarzua2017,Rajegowda2018analysing,Fu2018} or graphene \cite{Fu2017}, together with novel experiments by Bregulla et al. \cite{Bregulla2016}, which first reported a microscale manifestation of thermo-osmotic flows. Thermo-osmotic flows could in particular be boosted by the slip boundary condition (BC), which describes the velocity jump $\vs$ at the interface by a general expression first proposed by Navier \cite{Navier1823,Cross2018}:
 \begin{equation}
    \vs = b \frac{\partial v}{\partial z} \Bigg|_{z = \zs}, 
    \label{eq:slipBC_thermoosmosis}
\end{equation}
where $\zs$ corresponds to the shear plane position \cite{Herrero2019} and $b$ is the slip length \cite{Bocquet2007}. The role of interfacial hydrodynamics for thermo-osmosis has already been explored in the literature \cite{Ganti2017,Fu2017,Wang2020net}. Furthermore, in recent work on thermo-electricity, the critical role of the solvent enthalpy in describing the response has been highlighted for a modelled, highly hydrophobic surface \cite{Fu2019}. 

Following this work, we propose in this communication an analytical framework with the objective to predict thermo-osmosis of aqueous electrolytes confined by charged surfaces, extendable to thermoelectricity and thermophoresis. The solvent contribution and the electrostatic ionic interactions are shown to play the leading role along with hydrodynamic slip.
We apply the model to a wide range of systems, varying the wetting interaction, salt type and concentration, and the surface charge. We report large thermo-osmotic responses, comparable to the highest responses predicted for special systems as inferred from previous simulations \cite{Ganti2017,Fu2017,Oyarzua2017,Fu2018}, as well as a change of sign in the flow direction. Such change of sign cannot be predicted by only considering electrostatic interactions, and can be crucial in order to interpret the different experimental results reported in the literature.

\begin{figure}
\begin{minipage}{0.45\linewidth}%{0.24\textwidth}
\begin{subfigure}{0.99\textwidth}
    \centering
    \includegraphics[width=0.95\linewidth]{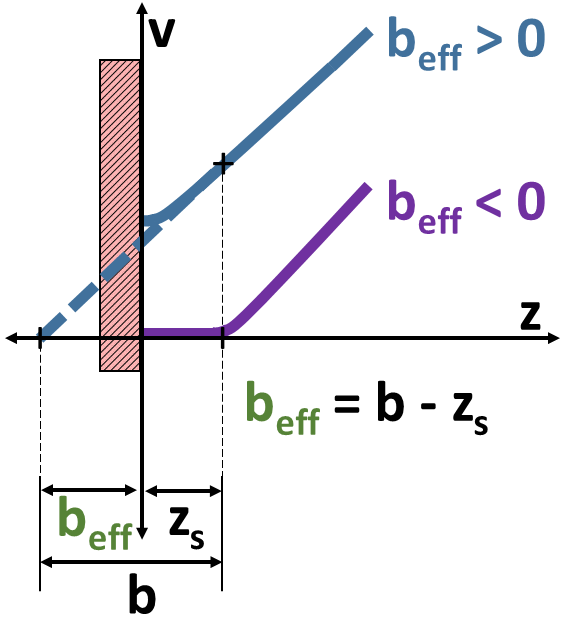}
    \caption{}
    \label{fig:HBC_beff}
\end{subfigure}\\%[\baselineskip]
\begin{subfigure}[b]{\linewidth}
	\centering
	\includegraphics[width=0.99\linewidth]{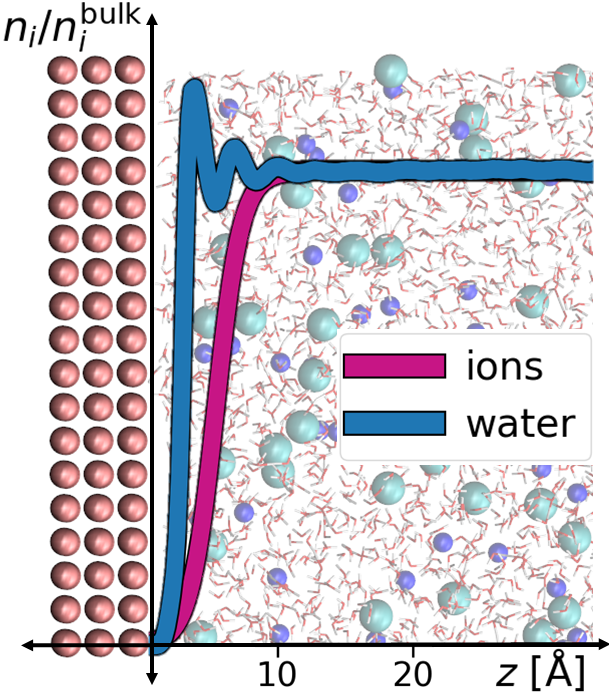}
    \caption{}
    \label{fig:system_density_v2}
\end{subfigure}
\end{minipage}
\begin{subfigure}{0.52\linewidth}
    \centering
    \includegraphics[width=0.99\linewidth]{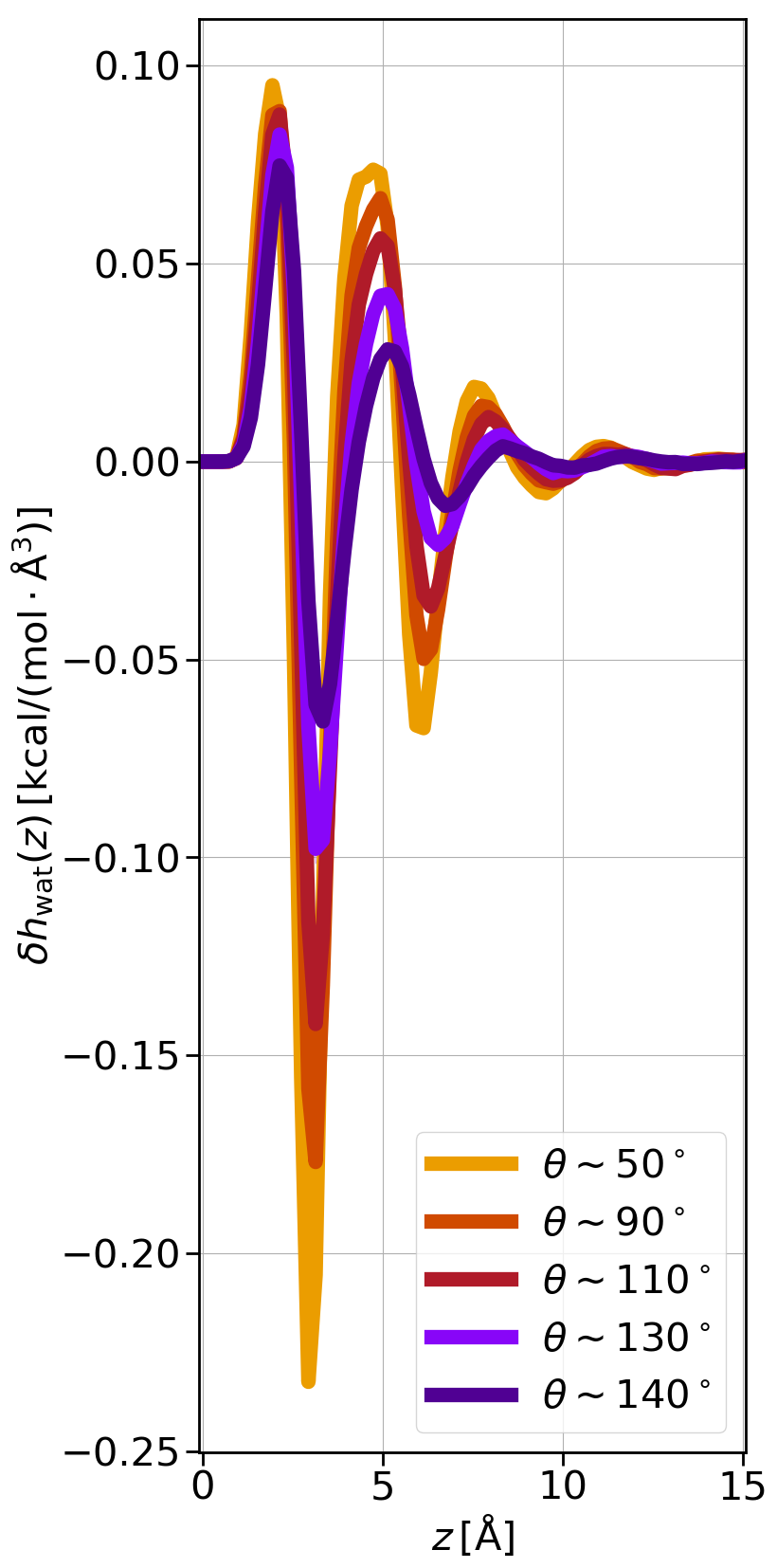}
    \caption{}
    \label{fig:dhw_eps_v2}
\end{subfigure}
\caption{(a) Schematics of the effective slip length $\beff$ as a function of the slip length $b$ and the shear plane position $\zs$. We distinguish between the slip situation ($\beff > 0$) and the stagnant layer situation ($\beff < 0$). (b) Modelled system for the measures of water enthalpy excess density and slip, together with the normalized density profiles of water and ions (Na$^+$ and Cl$^-$ in the picture), with $z$ the distance to the bottom wall. (c) Water enthalpy excess density $\delta h_\mathrm{wat}$ profiles, with $z$ the distance to the bottom wall, for different wetting angles $\theta$, controlled by the interaction energy between the liquid and the solid atoms $\eps_\mathrm{LS}$.}
\label{fig:fig1}
\end{figure}

\section*{Theoretical framework}
The thermo-osmotic response of a liquid-solid interface is quantified by the thermo-osmotic coefficient $M_\tosm$, defined from the relation $v_{\tosm} = M_\tosm (-\nabla T/T)$, where $\nabla T/T$ is the relative temperature gradient parallel to the wall, and $v_\tosm$ is the generated thermo-osmotic velocity far from the interface \cite{Anderson1989}.
In Ref.~\citenum{Fu2017}, the authors propose a modification to the classical Derjaguin theory \cite{Derjaguin1987} and show that, in order to take into account the hydrodynamic BC, the thermo-osmotic response coefficient can be expressed as (see the ESI):
\begin{equation}
    M_\tosm = \frac{1}{\eta} \int_{z_0}^\infty (z - \zs + b) \delta h(z) \dd z, 
    \label{eq:Mto}
\end{equation}
where $\eta$ is the liquid viscosity, $z$ the distance to the surface, $\delta h$ the enthalpy excess density, $b$ the slip length --defined from Eq.~\eqref{eq:slipBC_thermoosmosis}, and $\zs$ corresponds to the shear plane position. One can account for the presence of slip or a stagnant layer close to the wall by introducing an effective slip length $\beff = b - \zs$, see Fig.~\ref{fig:HBC_beff}. When $\beff \ge 0$ (slip situation), the velocity profile does not vanish in the water slab and therefore the integral in Eq.~\eqref{eq:Mto} should be performed from the wall position considered at zero, $z_0 = 0$. If on the contrary $\beff < 0$ (stagnant layer situation), then $\beff$ identifies with the size of a stagnant layer present at the liquid-solid interface, where the liquid velocity vanishes. In this case the stagnant layer does not contribute to the integral Eq.~\eqref{eq:Mto} and consequently $z_0 = -\beff$.

With regard to the enthalpy excess density $\delta h$, the standard approach \cite{Ruckenstein1981,Derjaguin1987} assumes that it is mostly determined by the electrostatic enthalpy of ions in the EDL, $\delta h_\mathrm{el}(z) = \rhoe(z) V(z) + p(z)$, where  $\rhoe$ is the charge density, $V$ is the local electric potential and $p$ is the pressure. Using the Poisson equation $\rhoe = -\eps \dd^2_z V$ (assuming a constant solvent permittivity $\eps$) and considering mechanical equilibrium along the $z$ direction, $\frac{\dd p}{\dd z} = -\rhoe \frac{\dd V}{\dd z}$, $\delta h_\mathrm{el}$ is then usually expressed in terms of the electrostatic potential as:
\begin{equation}
    \delta h_\mathrm{el}(z) = -\eps V(z) \frac{\dd^2V}{\dd z^2} + \frac{\eps}{2}\qty( \frac{\dd V}{\dd z} )^2.
    \label{eq:dhel}
\end{equation} 
Accordingly, $\delta h_\mathrm{el}$ vanishes outside the EDL, whose thickness is given by the Debye length $\debye$, controlled by the salt concentration \cite{Markovich2016charged}. 
The corresponding contribution to the thermo-osmotic response, $M_\tosm^\mathrm{el}$, can be then computed analytically within the mean-field Poisson-Boltzmann framework \cite{Herrero2021PB} (see the ESI).

Although the model proposed by Derjaguin et al. can predict $M_\tosm$ experimental orders of magnitude under certain conditions \cite{Bregulla2016}, it fails to describe the amplitude of the responses predicted in the literature \cite{Ganti2017,Fu2017,Oyarzua2017,Fu2018}, the thermo-osmotic response reported for pure water in uncharged membranes \cite{Mengual1978}, as well as the experimental discrepancies observed in $M_\tosm$ sign \cite{Derjaguin1980,Rusconi2004,Nedev2015,Bregulla2016}.
Aside of the electrostatic ionic interactions, other contributions to $\delta h$ can be important. Such contributions are related to the solvent (water in the present work), the ion solvation, and the water dipole orientation in the electric double layer. After comparing all the different contributions to $M_\tosm$ (see the ESI), the two main ones are (in the case of symmetric salts such as NaCl or KCl):
\begin{equation}
    M_\tosm \simeq M_\tosm^\mathrm{wat} + M_\tosm^\mathrm{el*} , 
    \label{eq:Mto_finaldecomposition}
\end{equation}
related to the solvent enthalpy excess density $\delta h_\mathrm{wat}$, and to a modified electrostatic term $\delta h_\mathrm{el}^*$, accounting for the depletion of the ions in the vicinity of the wall (see density profiles in Fig.~\ref{fig:system_density_v2}). 

Defining the characteristic depletion length as $\dl$, one can efficiently account for this effect by imposing a vanishing potential in the interfacial region where there are no ions. This defines, for a semi-infinite channel, $\delta h_\mathrm{el}^* = \delta h_\mathrm{el}$ for $z>\dl$ and $\delta h_\mathrm{el}^*=0$ otherwise. With regard to the water contribution, $\delta h_\mathrm{wat}$ can be computed as the sum of the different atomic contributions, $\delta h_\mathrm{wat} = \delta h_\mathrm{O}+\delta h_\mathrm{H}$, where the atomic enthalpy density for an element $i$ is defined as:
\begin{equation}
    \delta h_i(z) = \qty[\delta u_i(z) + \delta p_i (z) ]n_i(z),
\end{equation}
where $\delta \mathcal{A}(z) = \mathcal{A}(z) - \mathcal{A}_\mathrm{bulk}$; with $\mathcal{A}_\mathrm{bulk}$ the bulk value of the physical property $\mathcal{A}$, $u_i$ the energy per atom, $p_i$ the stress per atom (a practical difficulty with measuring this term is discussed in the ESI), and $n_i$ the atomic number density profile. 

To compute the solvent term $\delta h_\mathrm{wat}$ and the hydrodynamic BC as a function of wetting, we ran MD simulations using the LAMMPS package \cite{Plimpton1995}. The system consisted of an aqueous electrolyte (constituted by SPC/E water \cite{Berendsen1987} and NaCl, such as the bulk salt concentration was $\nbulk \sim 1\,$M, following Ref.~\citenum{Huang2007}) confined between generic uncharged Lennard-Jones (LJ) walls and graphene; see Fig.~\ref{fig:system_density_v2} and details in the ESI. We confirmed that in the case of symmetric salts the solute enthalpy, even at large concentrations, does not affect the total enthalpy profile, which is controlled by the solvent.

The solid wall atoms were frozen and the oxygen-solid (LS) interactions were varied between the hydrophobic (with contact angle $\theta \sim 140^\circ$) and hydrophilic ($\theta \sim 50^\circ$) values given in Ref.~\citenum{Huang2007} for LJ walls (the values for $\theta$ and $\beff$ can be found in the ESI). In Fig.~\ref{fig:dhw_eps_v2} one can observe the typical shape of $\delta h_\mathrm{wat}$ profile for different wettings. We note that the most hydrophilic situation ($\theta \sim 50^\circ$) is considered as a no-slip situation with $b=0.0~$\AA{} corresponding to a stagnant layer ($\beff < 0$). For simplicity, we also did not take into account in our model the coupling between the surface charge and slip \cite{Xie2020}. Using the proposed analytical framework, we explored a range of experimentally accessible values for the surface charge density $\Sigma$ and the salt concentration $\nbulk$: $\Sigma$ was varied between $-1$ and $-300\,$mC/m$^2$, and $\nbulk \in \{10^{-4}, 1\}~$M corresponding to a Debye length $\debye \in \{0.3,30\}~$nm. 

The objective of this communication is to present a general simple model, and with that regard some approximations are applied in order to explore a broad range of parameters. Nevertheless, the validity of the approximations we use is consistent with the range of parameters we explored, such as the choice of a lower boundary for $\debye$ comparable to the size of water's first absorption layer (where water solvent properties should be accounted for in the calculations and solvation and water properties should not be considered separately), as well as the upper boundary for $\Sigma$, under which the mean-field Poisson-Boltzmann descriptions should remain valid \cite{Herrero2021PB} (see the ESI).

\section*{Results and Discussion}
\begin{figure*}
 \centering
 \includegraphics[width=0.99\textwidth]{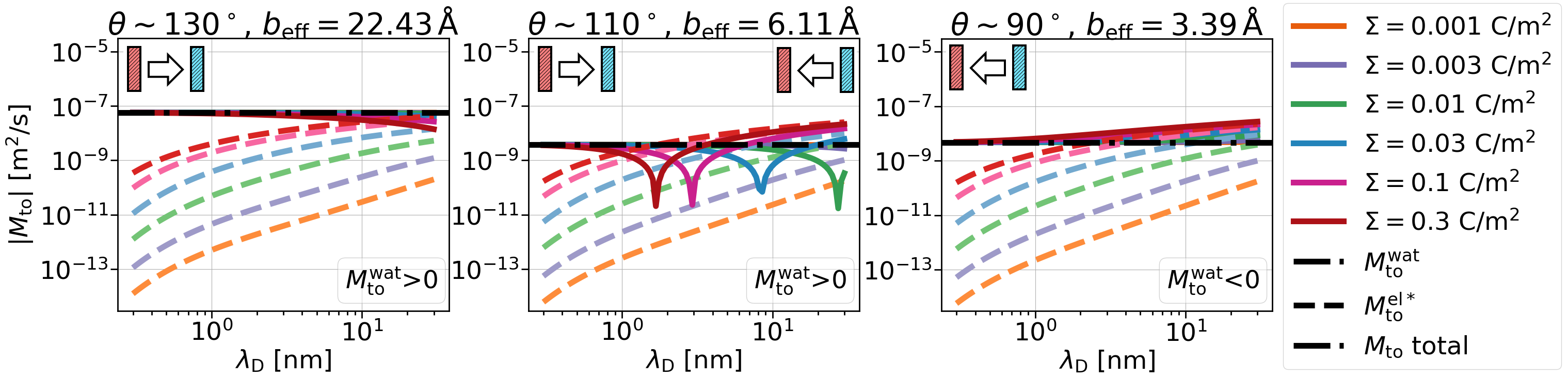}
 \caption{Thermo-osmotic response coefficient $M_\tosm$ (solid lines) as a function of the Debye length, for different wettings and surface charges. In all the graphs the two main contributions, water $M_\tosm^\mathrm{wat}$ (dash-dotted lines) and modified electrostatic $M_\tosm^\mathrm{el*}$ (dashed lines), are also represented. While $M_\tosm^\mathrm{el*}$ is always negative, the sign of $M_\tosm^\mathrm{wat}$ depends on wetting, so that the total response can be thermophilic or thermophobic depending on wetting, surface charge and Debye length.}
 \label{fig:totalMto}
\end{figure*}

From Eq.~\eqref{eq:Mto_finaldecomposition}, we expect that $M_\tosm$ is controlled by the competition between water and electrostatic contributions, depending on wetting, $\Sigma$ and $\nbulk$ (or analogously $\debye$). In Fig.~\ref{fig:totalMto} the total thermo-osmotic response is represented for all the wettings considered, together with $M_\tosm^\mathrm{wat}$ (independent of $\debye$ and $\Sigma$) and $M_\tosm^\mathrm{el*}$ (which presents a very small wetting effect). We observe in this figure the rich behavior resulting from that competition, where the water term mostly dominates for the most hydrophobic surfaces ($\theta \gtrsim 110^\circ$), while for the most hydrophilic surfaces ($\theta \lesssim 110^\circ$) electrostatic can dominate for the larger $\debye$. We can also see a large variation of $M_\tosm$ for the different wettings, ranging from $10^{-9}$ to $10^{-7}~$m$^2$/s for the most hydrophobic case.

A striking result from Fig.~\ref{fig:totalMto} is the transition for intermediate wettings between a thermophobic behavior ($M_\tosm > 0$) at high salt concentration (small $\debye$) to a thermophilic behavior ($M_\tosm < 0$) at low salt concentration, see for instance $\theta \sim 110^\circ$. In agreement with previous predictions \cite{Bregulla2016}, the electrostatic contribution $M_\tosm^\mathrm{el*}<0$, yielding thermophilic behavior independently of the sign of the surface charge. In contrast, the water term exhibits a change of sign when varying the wetting (see the ESI). Such change of sign of $M_\tosm^\mathrm{wat}$ happens at $\theta \sim 110^\circ$ and thus, for $\theta \gtrsim 110^\circ$, $\sgn(M_\tosm^\mathrm{wat}) = - \sgn(M_\tosm^\mathrm{el*})$ resulting in a change of sign of $M_\tosm$ when $\debye$ is such that  $\abs{M_\tosm^\mathrm{wat}} = \abs{M_\tosm^\mathrm{el*}}$. Although this change of behavior happens for all $\theta \gtrsim 110^\circ$, for the most hydrophobic cases it takes place for $\debye$ values higher than the ones considered in this study and far from the limits of validity of the Poisson-Boltzmann framework considered in the computation of $M_\tosm^\mathrm{el^*}$. Even so, within our parameters range, we can still observe for $\theta \sim 130^\circ$ a decrease of the total response for high $\debye$, which goes against the standard expectation and can only happen if water and electrostatic contributions have opposite signs. In contrast, for the most hydrophilic cases (\emph{e.g.} $\theta \sim 90^\circ$), $\sgn(M_\tosm^\mathrm{wat}) = \sgn(M_\tosm^\mathrm{el*})$ and $M_\tosm$ does not change sign for any $\debye$ value.

It is interesting to note that a similar change of sign has been found in the context of thermophoresis experiments \cite{Gaeta1982,Putnam2005,Wurger2008}. This change of sign is commonly attributed to the so-called thermopotential $\psi_0$ \cite{Wurger2010}. Such thermopotential appears for ions having an asymmetric mobility, from imposing no ionic flux conditions, and it generates an electro-osmotic flow, which can go against the thermo-osmotic flow and reverse the total flow direction. Nevertheless, $\psi_0$ should disappear by allowing ionic fluxes through the channel, and as a consequence the change of sign would disappear. By introducing the water contribution to the thermo-osmotic response, we propose a more fundamental understanding of such change of sign, which should persist independently of the boundary conditions on the fluxes through the channel.

\begin{figure}
\centering
  \includegraphics[width=0.95\linewidth]{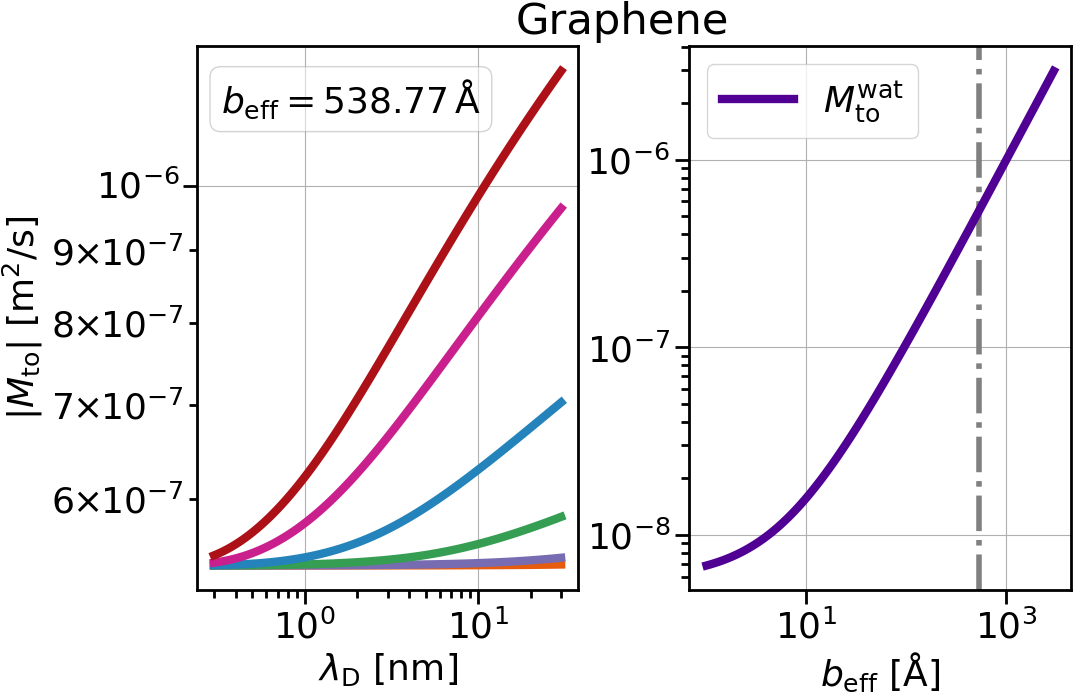}
  \caption{Thermo-osmotic response coefficient for graphene walls. Left: total response as a function of the Debye length, for different surface charges as in Fig.~\ref{fig:totalMto}. Right: water contribution $M_\tosm^\mathrm{wat}$ as a function of the effective slip length. The dash-dotted line indicates the $\beff$ value obtained from MD simulations. For graphene walls, $M_\tosm^\mathrm{wat}$ and $M_\tosm$ are always negative, corresponding to a thermophilic flow.}
  \label{fig:Mto_GR}
\end{figure}

The proposed $M_\tosm^\mathrm{el*}$ and $M_\tosm^\mathrm{wat}$ decomposition allows us to obtain agreement of the order of magnitude with the experimental results of $M_\tosm$ \cite{Bregulla2016}, on the order of $10^{-10}-10^{-9}~$m$^2$/s. Such agreement is specially significant for hydrophilic surfaces and in the stagnant layer situation (see the ESI), typical of experiments due to the presence of imperfections in the solid surface, when $M_\tosm^\mathrm{wat}$ decreases and $M_\tosm^\mathrm{el*}$ may dominate for a broader range of Debye lengths. Because $M_\tosm^\mathrm{wat}$ increases when increasing the slip, one interesting surface is the one constituted by graphene, with an effective slip length of $\beff =538.77\,$\AA{}, which we obtained in MD simulations for NaCl aqueous solution at room temperature (see the ESI). In Fig.~\ref{fig:Mto_GR}~left one can observe a significant increase in both electrostatic and water $M_\tosm$ contributions, resulting in a large value of the total response $M_\tosm \sim 10^{-6}~$m$^2$/s for this interface. Because $M_\tosm^\mathrm{el}$ does not vary significantly with wetting and the order of magnitude of the total response is given by the water contribution, one can expect $M_\tosm \sim M_\tosm^\mathrm{wat}$ for graphene. In Fig.~\ref{fig:Mto_GR}~right one can see how $M_\tosm^\mathrm{wat}$ is affected by the effective slip. In this figure one observes that a large $M_\tosm$ value may be obtained for very slipping systems (as CNT, where slip values of $b \sim 300~$nm have been reported at room temperature for tube radius of $R \sim 15~$nm \cite{Secchi2016}), although it is important to note that the presence of a stagnant layer or defects in the surface (resulting in smaller $\beff$) may decrease the large predicted thermo-osmotic response, down to $10^{-9}~$m$^2$/s.

\section*{Conclusions}
We proposed here an analytical framework aimed at predicting the thermo-osmotic response of aqueous electrolytes for a wide range of nanofluidic systems and experimental conditions. While the standard picture relates the response to the ion electrostatic enthalpy in the electrical double layer close to charged walls, we show first that this contribution to the interfacial enthalpy may be negligible when compared to the water contribution for a broad range of parameters, and second that it should be slightly lowered due to the depletion of ions from the solid surface.

The competition between the modified electrostatic and water contributions and the impact of the hydrodynamic boundary condition leads to a rich phenomenology that we illustrated here. First, our theory predicts a higher thermo-osmotic response at low $\debye$ than the one expected from only considering the electrostatic contribution. Second, the proposed model also predicts a transition between a thermophobic behavior at low salt concentrations to a thermophilic behavior at high salt concentrations for intermediate wettings. Such transition has also been observed in thermophoresis experiments \cite{Gaeta1982,Putnam2005,Wurger2008}, and is commonly attributed to the existence of a thermopotential which is, however, limited to particular boundary conditions imposing no ionic fluxes in the bulk liquid. In contrast, our interpretation of the change of sign is more general and independent on the nanofluidic channel boundary conditions, and opens the way to manipulate thermally induced nanoscale flows with a pinch of salt. Third, we predict intense thermally induced flows for slipping systems, with orders of magnitude comparable to the ones reported from MD simulations of water thermo-osmosis in CNTs \cite{Oyarzua2017,Fu2018} or on uncharged planar walls \cite{Ganti2017,Fu2017}. Therefore, our analysis predicts that very strong thermo-osmotic flows can be obtained not only for special systems such as carbon nanotubes, but also with more common hydrophobic charged surfaces, paving the way to explore other common and affordable charged surfaces ensuring the absence of a liquid stagnant layer at the interface so that slip can boost the response. 

The importance of the solvent contribution in thermo-osmosis of aqueous electrolytes, together with a modification of the classical electrostatic term, opens the way to several perspectives.
With respect to the electrostatic term, a more accurate description of thermo-osmosis should take into account spatial heterogenities of the dielectric and viscosity profiles at the interface \cite{Hoang2012localviscosity,Bonthuis2013,rezaei2021interfacial}. Also, for very asymmetric salts, such as NaI, the ion-size-dependent hydrophobic solvation term should be considered, \emph{e.g.} through the modified Poisson-Boltzmann framework described in Refs.~\citenum{Huang2007,Huang2008}. 
Regarding the water term, it is left to determine the impact of the surface charge and its distribution on water contribution to the response. Besides, one should take into account the limits of considering pure water simulations as an approximation of the water enthalpy contribution. For high concentrations, steric effects should be accounted for, and ions can affect water viscosity \cite{Kim2012}. Nevertheless such effects correspond to extreme $\nbulk$ values \cite{Herrero2021PB} and they should not understate one of the main results of the present chapter: the great $M_\tosm $ value found for slipping surfaces.
Finally, it is straightforward to extend the current model to predict the thermoelectric \cite{Hartel2015,Dietzel2016,Jin2021} and thermodiffusive \cite{DiLecce2020} response, with promising applications for electricity production from waste heat or to refine large-scale continuum descriptions \cite{Dietzel2017}. Overall, our predictions call for future experimental verification, and could be exploited for the design of innovative solutions for heat harvesting applications.

\begin{acknowledgments}
The authors thank Aymeric Allemand, Anne-Laure Biance, Li Fu and Christophe Ybert for fruitful discussions. 
We are also grateful for HPC resources
from GENCI/TGCC (grants A0070810637 and A0090810637), and 
from the PSMN mesocenter in Lyon. 
This work is supported by the ANR, Project ANR-16-CE06-0004-01 NECtAR. LJ is supported by the Institut Universitaire de France.
\end{acknowledgments}

%\bibliography{biblio_thesis.bib,biblio_manual.bib}

\end{document}